\documentclass[a4paper,numcites,final]{aipproc}
\setcitestyle{numbers}
\layoutstyle{8x11double}

\newcommand\FP{Faddeev--Popov\ }
\newcommand\YM{Yang--Mills\ }
\newcommand\oh{{\textstyle\frac{1}{2}}}

\graphicspath{{figures/}}
\begin{document}

\title{Properties of the approximate Yang--Mills ground-state\\wave functional in $2+1$ dimensions\thanks{Contribution presented by \v S.\ Olejn\'\i k. This research was supported in part by the U.S.\ DOE under Grant No.\ DE-FG03-92ER40711 (J.G.), and by the Slovak Grant Agency for Science, Project VEGA No.\ 2/0070/09, by ERDF OP R\&D, Project CE QUTE ITMS~26240120009, and via CE SAS QUTE (\v{S}.O.)
}}


\author{Jeff Greensite}{
  address={Physics and Astronomy Dept., San Francisco State University, San Francisco, CA 94132, USA}
}

\author{\v Stefan Olejn\'\i k}{
  address={Institute of Physics, Slovak Academy of Sciences, D\'ubravsk\'a cesta 9, SK--845 11 Bratislava, Slovakia}
}

\begin{abstract}
We review properties and lattice evidence in support of the recently proposed temporal-gauge Yang--Mills vacuum wave functional in $2+1$ dimensions.
\end{abstract}

\maketitle


\section{Introduction}

	Confinement is wired in the structure of the ground state of the Yang--Mills (YM) theory. In the Schr\"odinger picture in a physical gauge, the information is carried by the vacuum wave functional. One can approach the problem from various angles:
%

1.~Try to develop a systematic strong-coupling expansion of the vacuum wave functional (VWF) \cite{Greensite:1979ha}.

2.~Express the theory in cleverly selected variables and look for the VWF in terms of these variables \cite{Karabali:1998yq}.

3.~Make a variational Ansatz for the wave functional in a certain gauge and find its parameters by minimizing (e.g.) the expectation value of the YM Hamiltonian \cite{HR-AS}.

4.~Guess an approximate form of the wave functional and test its consequences \cite{Samuel:1996bt,Greensite:2007ij}.
%

	We follow the last avenue: we have proposed a guess of a simple vacuum wave functional in the temporal gauge \cite{Greensite:2007ij}, and will present here a few pieces of evidence in favor of the claim that this simple form is not far from the true vacuum wave functional of the Yang--Mills theory.

	In the hamiltonian formulation in $D=2+1$ dimensions and temporal gauge, the vacuum wave functional satisfies the Schr\"odinger equation:
\begin{equation}
\displaystyle\int d^2 x\left(-{\textstyle\frac{1}{2}}
\frac{\delta^2}{\delta A_k^a(x)^2}+{\textstyle\frac{1}{4}}B^a(x)^2\right)\Psi_0[A]=E_0\Psi_0[A]
\end{equation}
together with the Gauss-law constraint:
\begin{equation}
\left(\delta^{ac}\partial_k+
g\epsilon^{abc}A_k^b\right)
\frac{\delta}{\delta A_k^c}\Psi_0[A]=0.
\end{equation}

	At large distance scales one expects effectively:
\begin{equation}
\Psi_0^{\mathrm{eff}}[A]\approx\exp\left[-\mu\int d^2x\; B^a(x)B^a(x)\right].
\end{equation}
This form has the property of \emph{dimensional reduction} \cite{Greensite:1979yn}: The computation of a spacelike loop in \linebreak $2+1$ dimensions reduces to the calculation of a Wilson loop in \YM theory in $2$ euclidean dimensions, which is known to exhibit the area law. However, the true VWF cannot be that simple, as it leads to incorrect physical predictions at short and intermediate distances.

\section{A guess at an approximate VWF}

	The starting point for our guesswork [for SU(2) gauge theory] has been the QED vacuum wave functional \cite{Wheeler}, whose simplest generalization reads
\begin{equation}\label{GO1}
\Psi_0[A]=\exp\left[-\oh\int d^2x d^2y\; B^a(x)V^{ab}(x,y;g)B^b(y)\right],
\end{equation}
where $V^{ab}$ is an adjoint operator which fulfills, for vanishing coupling, 
\begin{equation}
\displaystyle\lim_{g\to 0} V^{ab}(x,y;g)=\left(\frac{\delta^{ab}}{\sqrt{-\nabla^2}}\right)_{xy}.
\end{equation}
We choose
\begin{equation}\label{GO2}
V^{ab}(x,y;g)=\left(\frac{1}{\sqrt{-{\cal D}^2-
\lambda_0+m^2}}\right)_{xy}^{ab},
\end{equation}
where ${\cal D}_k[A;g]$ is the covariant derivative in the adjoint representation, ${\cal D}^2={\cal D}_k\cdot{\cal D}_k$ the adjoint covariant laplacian, $\lambda_0$ is the lowest eigenvalue of $(-{\cal D}^2)$, and $m$ is a constant (mass) parameter proportional to $g^2\sim 1/\beta$. The expression for the VWF is written in the continuum notation, but assumed to be properly defined on a lattice.

	In Ref.\ \cite{Greensite:2007ij} we have provided a series of analytical arguments in favor of the proposed form:
%

1.~By construction, $\Psi_0[A]$ becomes the VWF of electrodynamics in the free-field limit ($g\to0$).

2.~It is a good approximation to the true vacuum also for strong fields constant in space, varying only in time.

3.~The part of the VWF that depends on slowly varying fields $B_\mathrm{slow}$ takes on the dimensional-reduction form. The fundamental string tension, at a given $\beta$, is then easily computed as $\sigma_\mathrm{F} = 3m/4\beta$.

4.~If one takes the mass $m$ in the VWF as a free variational parameter and computes (approximately) the expectation value of the YM hamiltonian, one finds that a non-zero (finite) value of $m$ is energetically preferred.
%

\section{Numerical evidence}

\begin{figure}
\centerline{\includegraphics[width=\columnwidth]{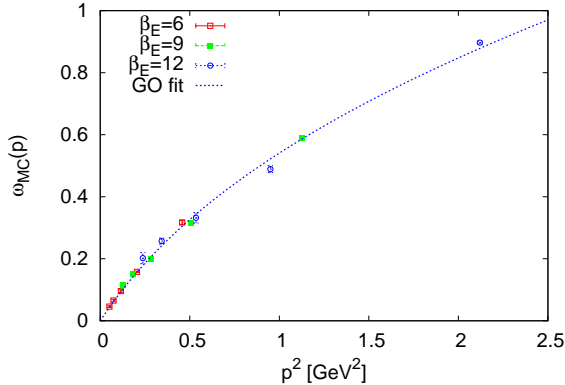}}
\caption{Cumulative data for $\omega_{\mathrm{MC}}$ vs.\ $p^2$ in physical units, on lattices of extensions
$L=16,24,32,40,48$, and euclidean lattice couplings $\beta_E=6,9,12$.  The curve represents $\omega_{\mathrm{GO}}(p)$ using the parameters of $m$ and $g^2$ quoted in the text.}
\label{om}
\end{figure}

	 To estimate how good or bad the proposed approximate vacuum state is, we have also performed some numerical tests. We compare a set of quantities computed in two ensembles of lattice configurations:
	%

1.~\emph{``Recursion'' lattices} --- independent 2-d lattice configurations generated with the probability distribution given by the proposed VWF, for a choice of its parameters $m$ and $\beta=4/g^2$. The recursion method was proposed and described in detail in Ref.\ \cite{Greensite:2007ij}. 

2.~\emph{Monte Carlo lattices} --- 2-d slices of configurations generated by MC simulations of the 3-d euclidean SU(2) lattice gauge theory with the standard Wilson action with coupling $\beta_E$; from each configuration, only one (random) slice at fixed euclidean time was taken. 

	To compare results for these two ensembles, one needs a way of fixing parameters $(m,\beta)$ of the VWF that correspond to the Wilson-action coupling $\beta_E$ of the MC ensemble at a fixed lattice size $L$. The simplest way, used in Refs.~\cite{Greensite:2007ij,Greensite:2010tm}, is to choose $\beta=\beta_E$, and fix $m$ at given $\beta_E$ and $L$ to get the correct value of the fundamental string tension $\sigma_\mathrm{F}(\beta_E,L)$. (This will be called \emph{variant A} below.) With this choice we computed the equal-time connected $B^2$--$B^2$ correlator and determined the value of the mass gap from a best fit to its exponential fall-off at large distances. The result for recursion lattices was compared to the values of the $0^+$ glueball mass computed in simulations of the 3-d YM theory by Meyer and Teper \cite{Meyer:2003wx}. The deviations were at the level of at most 6\%.

\begin{table}[b!]
\begin{tabular}{ c c  c  c c  c c }
\hline
& & & \multicolumn{2}{c}{\emph{variant A}} & \multicolumn{2}{c}{\emph{variant B}}\\
\cline{4-5}\cline{6-7}
${\beta_\mathrm{E}}$ & ${L}$ & ${a(\beta_\mathrm{E},L)}$ & ${\beta}$  & ${m}$ & ${\beta}$ & ${m}$ \\\hline
6 & 24 & 0.577 & 6 & 0.515 & 4.734 & 0.445\\
9 & 32 & 0.367 & 9 & 0.313 & 7.434 & 0.283\\\hline
\end{tabular}
\caption{Parameters and lattice sizes.}
\label{tab1}
\end{table}

	Another possibility (\emph{variant B}) is to find $(m,\beta)$ corresponding to a given $(L,\beta_E)$ by using numerically determined values of the true VWF for some trial gauge-field configurations (non-abelian constant fields, abelian or non-abelian plane waves). The square of the VWF can be computed numerically in simulations of the 3-d YM theory by the method proposed long ago by Greensite and Iwasaki \cite{Greensite:1988rr}. Details of the method and results for various types of configurations will be presented elsewhere \cite{Greensite:in-prep}. We will only use results for a set of abelian plane waves with the longest wavelength at a given lattice size $L$, $\lambda=L$, and varying amplitudes:
\begin{eqnarray}
&&U_1^{(j)}=\sqrt{1-a_j(n_2)^2}\mathbf{1}_2+i a_j(n_2)\sigma_3,\quad
U_2^{(j)}=\mathbf{1}_2,\nonumber\\
&&a_j(n_2)=\sqrt{\frac{\alpha+\gamma j}{L^2}}\cos\frac{2\pi n_2}{L}.
\end{eqnarray}
The measured squared VWF at a given $\beta_E$ can be parametrized as $\vert\Psi_\mathrm{MC}[U^{(j)}]\vert^2\propto\exp(-R_\mathrm{MC}[U^{(j)}])$ with: 
\begin{equation}
R_\mathrm{MC}[U^{(j)}]=2(\alpha+\gamma j)\omega_\mathrm{MC}(p)+\mbox{ const}
\end{equation}
with
{$p^2=2\left(1-\cos\frac{2\pi}{L}\right)$}, while for our proposed wave functional
\begin{equation}
{R_\mathrm{GO}[U^{(j)}]=2(\alpha+\gamma j)\omega_\mathrm{GO}(p)+\mbox{ const}},
\end{equation}
\begin{equation}\label{omega-GO}
{\omega_\mathrm{GO}(p)=\frac{1}{g^2}\frac{p^2}{\sqrt{p^2+m^2}}}.
\end{equation}
Values of $(m,g)$ can be determined by the best fit of the data for $\omega_\mathrm{MC}(p)$ at a given $(L,\beta_E)$  by the function~(\ref{omega-GO}). The quality of the fit is illustrated in Fig.\ \ref{om}, the parameters of our wave functional in physical units come out to be $g^2_\mathrm{phys}=1.465, m_\mathrm{phys}=0.771$. In lattice units we then use: 
\begin{equation}
{{a(\beta_\mathrm{E},L)}={\sqrt{{\sigma_\mathrm{f}(\beta_\mathrm{E},L)}}}/{0.44\ \mathrm{GeV}}},
\end{equation}
$$
{{m}(\beta_{\mathrm{E}},L)=m_\mathrm{phys}\;{a(\beta_\mathrm{E},L)}},\quad
{{g^2}(\beta_{\mathrm{E}},L)=g^2_\mathrm{phys}\;{a(\beta_\mathrm{E},L)}}.
$$
The sets of parameters used in further tests are summarized in Table \ref{tab1}.

	For both recursion and Monte Carlo lattices, we concentrated on a subset of quantities in the Coulomb gauge~\cite{Greensite:2010tm}. As argued by Gribov \cite{Gribov:1977wm} and Zwanziger \cite{Zwanziger:1998ez}, the low-lying spectrum of the \FP operator in Coulomb gauge probes properties of non-abelian gauge fields crucial for the confinement mechanism. The ghost propagator in Coulomb gauge and the color-Coulomb potential are directly related to the inverse of the \FP operator, and play a role in various confinement scenarios. In particular, the color-Coulomb potential represents an upper bound on the physical potential between a static quark and antiquark \cite{Zwanziger:2002sh}.

\begin{figure}
{\includegraphics[width=\columnwidth]{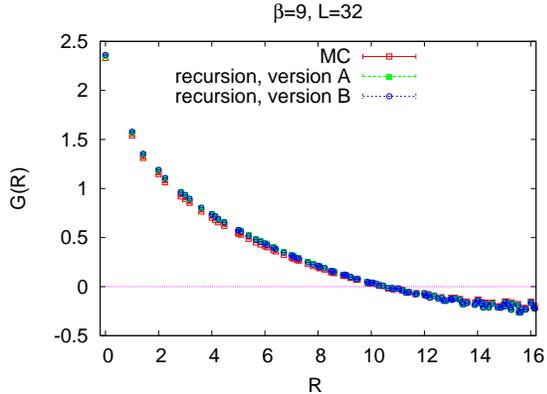}}
\caption{The Coulomb-gauge ghost propagator.}\label{fig:b9_l32_propagator}
\end{figure}

	Fig.\ \ref{fig:b9_l32_propagator} displays results for the ghost propagator in Coulomb gauge at $\beta=9$ on $32^2$ lattice. It was computed (in coordinate space) in the usual way from the inverse of the \FP operator (in the subspace orthogonal to trivial constant zero modes due to lattice periodicity). The agreement of the ghost propagator computed in both sets of lattices is almost perfect, for both variants of the choice of parameters of the approximate VWF.

\begin{figure}
{\includegraphics[width=\columnwidth]{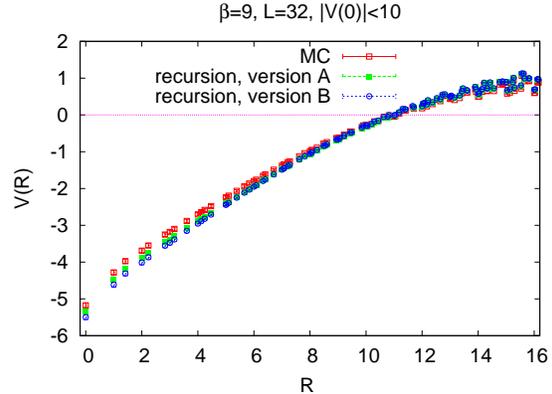}}
\caption{The color-Coulomb potential for $\kappa=10$.}\label{fig:b9_l32_potential}
\end{figure}

	The situation for the color-Coulomb potential is more complicated. There exist rare ``exceptional'' configurations with a very low (though still positive) lowest nontrivial eigenvalue of the \FP operator. These configurations were extremely difficult to gauge-fix to the Coulomb gauge. If one evaluates the potential in each single configuration, the exceptional ones possess a very high absolute value of the potential at the origin, $\vert V(0)\vert$. One can then classify configurations by their values of $\vert V(0)\vert$, and evaluate average potentials from sets of configurations satisfying a number of cuts $\vert V(0)\vert<\kappa$. Fig.\ \ref{fig:b9_l32_potential} shows results for $\kappa=10$ (satisfied by about 80\% lattices) at $\beta=9$ ($32^2$ lattice). The potentials agree quite well. However, the agreement between recursion and Monte Carlo lattices deteriorates with increasing $\kappa$. Still, the potentials stay almost identical for recursion lattices with both variant choices of VWF parameters.

\vspace*{2mm}
	\emph{In conclusion},	the proposed vacuum wave functional for the temporal-gauge SU(2) YM theory in 2+1 dimensions, Eqs.\ (\ref{GO1}, \ref{GO2}), is a fairly good approximation to the true ground state of the theory. We have accumulated analytical arguments and numerical results in its favor. The agreement of Coulomb-gauge quantities in recursion and Monte Carlo lattice ensembles is quite satisfactory for a bulk of the probability distribution, but there seems to be some disagreement in its tail.


%


\end{document}